\documentclass[conference]{IEEEtran}
\usepackage{cite}
\usepackage{graphicx,color}
\usepackage{amsmath,amssymb}
\usepackage{algorithmic}
\usepackage{array}
\usepackage[tight,footnotesize]{subfigure}
\usepackage{url}
\usepackage{amsthm}
\usepackage{enumerate}

\ifCLASSOPTIONcompsoc 
\usepackage[tight,normalsize,sf,SF]{subfigure}
\else 
\usepackage[tight,footnotesize]{subfigure}
\fi

\newcommand{\R}{\mathbb{R}}

\newcommand{\rank}{\mathrm{rank}}
\newcommand{\sgn}{\mathrm{sgn}}
\newcommand{\supp}{\mathrm{supp}}
\renewcommand{\P}{\mathcal{P}}
\newcommand{\PO}{\mathcal{P}_\Omega}
\newcommand{\PT}{\mathcal{P}_T}
\newcommand{\POp}{\mathcal{P}_{\Omega^\perp}}
\newcommand{\PTp}{\mathcal{P}_{T^\perp}}
\renewcommand{\S}{\mathbb{S}}
\renewcommand{\Pr}{\mathbb{P}}
\newcommand{\POj}{\mathcal{P}_{\Omega_j}}
\renewcommand{\mathbf}{\boldsymbol}
\newcommand{\event}{\mathcal{E}}

\renewcommand{\Re}{\mathbb{R}}

\theoremstyle{definition}
\newtheorem{thm}{Theorem}

\theoremstyle{definition}
\newtheorem{lemma}{Lemma}

\theoremstyle{definition}

\begin{document}
%
\title{Dense Error Correction for Low-Rank Matrices via Principal Component Pursuit}

\author{\IEEEauthorblockN{Arvind Ganesh\IEEEauthorrefmark{2},
John Wright\IEEEauthorrefmark{1},
Xiaodong Li\IEEEauthorrefmark{3}, 
Emmanuel J. Cand{\`{e}s}\IEEEauthorrefmark{3}$^,$\IEEEauthorrefmark{4} and
Yi Ma\IEEEauthorrefmark{1}$^,$\IEEEauthorrefmark{2}}
\IEEEauthorblockA{\IEEEauthorrefmark{1}Microsoft Research Asia, Beijing, P.R.C}
\IEEEauthorblockA{\IEEEauthorrefmark{2}Dept. of Electrical and Computer Engineering, UIUC, Urbana, IL 61801}
\IEEEauthorblockA{\IEEEauthorrefmark{3}Dept. of Mathematics, Stanford University, Stanford, CA 94305}
\IEEEauthorblockA{\IEEEauthorrefmark{4}Dept. of Statistics, Stanford University, Stanford, CA 94305}}


\maketitle

\begin{abstract}
We consider the problem of recovering a low-rank matrix when some of its entries, whose locations are not known a priori, are corrupted by errors of arbitrarily large magnitude. It has recently been shown that this problem can be solved efficiently and effectively by a convex program named Principal Component Pursuit (PCP), provided that the fraction of corrupted entries and the rank of the matrix are both sufficiently small. In this paper, we extend that result to show that the same convex program, with a slightly improved weighting parameter, exactly recovers the low-rank matrix even if ``almost all'' of its entries are arbitrarily corrupted, provided the signs of the errors are random. We corroborate our result with simulations on randomly generated matrices and errors.
\end{abstract}
\IEEEpeerreviewmaketitle

\section{Introduction} \label{sec:introduction}

Low-rank matrix recovery and approximation have been extensively studied lately for their great importance in theory and practice. Low-rank matrices arise in many real data analysis problems when the high-dimensional data of interest lie on a low-dimensional linear subspace. This model has been extensively and successfully used in many diverse areas, including face recognition \cite{Wright2009-PAMI}, system identification \cite{Fazel2004-ACC}, and information retrieval \cite{Papadimitriou2000-JCSS}, just to name a few. 

Principal Component Analysis (PCA) \cite{Jolliffe1986} is arguably the most popular algorithm to compute low-rank approximations to a high-dimensional data matrix. Essentially, PCA solves the following optimization problem:
\vspace{-0.08in}
\begin{equation}
\min_{L} \: \|D-L\| \quad \mathrm{s.t.} \quad \rank(L) \leq r,
\vspace{-0.06in}
\end{equation} 
where $D \in \R^{m \times n}$ is the given data matrix, and $\| \cdot\|$ denotes the matrix spectral norm. The optimal solution to the above problem is the best rank-$r$ approximation (in an $\ell^2$ sense) to $D$ \cite{Eckart1936-Psychometrika}.  Furthermore, PCA offers the optimal solution when the matrix $D$ is corrupted by i.i.d.~Gaussian noise. In addition to theoretical guarantees, the PCA can be computed stably and  efficiently via the Singular Value Decomposition (SVD). 

The major drawback of PCA is its brittleness to errors of large magnitude, even if such errors affect only a few entries of the matrix $D$. In fact, a single corrupted entry can throw the low-rank matrix $\hat L$ estimated by PCA arbitrarily far from the true solution. Unfortunately, these kinds of {\it non-Gaussian}, gross errors and corruptions are prevalent in modern data. For example, shadows in a face image corrupt only a small part of the image, but the corrupted pixels can be arbitrarily far from their true values in magnitude. 

Thus, the problem at hand is to recover a low-rank matrix $L_0$ (the principal components) from a corrupted data matrix 
$$
D = L_0+S_0,
$$ 
where the entries of $S_0$ can have arbitrary magnitude. Although this problem is intractable (NP-hard) to solve under general conditions, recent studies have discovered that certain convex program can effectively solve this problem under surprisingly broad conditions. The work of \cite{Candes2009-pp, Chandrasekaran2009-SYSID} has proposed a convex program to recover low-rank matrices when a fraction of their entries have been corrupted by errors of arbitrary magnitude {\it i.e.,} when the matrix $S_0$ is sufficiently sparse. This approach, dubbed Principal Component Pursuit (PCP) by \cite{Candes2009-pp}, suggests solving the following convex optimization problem:
\begin{equation}
\min_{L,S} \: \|L\|_* + \lambda\,\|S\|_1 \quad \mathrm{s.t.} \quad D =  L + S,
\label{eqn:pcp}
\end{equation}
where $\|\cdot\|_*$ and $\|\cdot\|_1$ denote the matrix nuclear norm (sum of singular values) and 1-norm (sum of absolute values of matrix entries), respectively, and $\lambda > 0$ is a weighting parameter. For square matrices of size $n \times n$, the main result of \cite{Candes2009-pp} can be summarized as follows:

\begin{quote}
{\em If the singular vectors of $L_0$ are not too coherent with the standard basis, and the support of $S_0$ is random, then solving the convex program \eqref{eqn:pcp} with $\lambda = n^{-1/2}$ exactly recovers $L_0$ of rank $O(n / \log^2 n)$ from errors $S_0$ affecting $\rho n^2$ of the entries, where $\rho > 0$ is a sufficiently small positive constant.}
\end{quote}

In this work, we extend the above result to show that under the same assumptions, \eqref{eqn:pcp} recovers low-rank matrices even if the fraction of corrupted entries $\rho$ is arbitrarily close to one, provided the signs of the errors are {\em random}. Equivalently speaking, {\it almost all} of the matrix entries can be badly corrupted by random errors. The analysis in this paper is a nontrivial modification to the arguments of \cite{Candes2009-pp} and leads to a better estimate of the weighting parameter $\lambda$ that enables this {\it dense error-correction} performance. We verify our result with simulations on randomly generated matrices. 


\section{Assumptions and Main Result} \label{sec:main}

For convenience of notation, we consider square matrices of size $n \times n$. The results stated here easily extend to non-square matrices.

\noindent{\bf Assumption A: Incoherence Model for $L_0$.} It is clear that for some low-rank and sparse pairs $(L_0,S_0)$, the problem of separating $M = L_0 + S_0$ into the components that generated it is not well-posed, e.g., if $L_0$ is itself a sparse matrix. In both matrix completion and matrix recovery, it has proved fruitful to restrict attention to matrices whose singular vectors are not aligned with the canonical basis. This can be formalized via the notion of {\em incoherence} introduced in \cite{Candes2008}. If $L_0 = U \Sigma V^*$ denotes a reduced singular value decomposition of $L_0$, with $U, V \in \R^{n \times r}$, and $\Sigma \in \R^{r \times r}$, then $L_0$ is $\mu$-incoherent if 
\begin{equation} \label{eqn:incoherence}
\left \{
\begin{array}{rcl}
\max_i\, \|U^*e_i\|^2 &\leq& \mu r / n,\\ \max_i\, \|V^*e_i\|^2 &\leq& \mu r / n,  \\
\|UV^*\|_\infty &\leq& \sqrt{\mu r / n^2},
\end{array}
\right .
\end{equation}
where the $e_i$'s are the canonical basis vectors in $\R^n$.  Here, $\|\cdot\|_\infty$ denotes the matrix $\infty$-norm (maximum absolute value of matrix entries).

\noindent{\bf Assumption B: Random Signs and Support for $S_0$.} Similarly, it is clear that for some very sparse patterns of corruption, exact recovery is not possible, e.g., if $S_0$ affects an entire row or column of the observation. In \cite{Candes2009-pp}, such ambiguities are avoided by placing a random model on $\Omega \doteq \mathrm{supp}(S_0)$, which we also adopt. In this model, each entry $(i,j)$ is included in $\Omega$ independently with probability $\rho$. We say $\Omega \sim \text{Ber}(\rho)$ whenever $\Omega$ is sampled from the above distribution. We further introduce a random model for the signs of $S_0$: we assume that for $(i,j) \in \Omega$, $\sgn((S_0)_{ij})$ is an independent random variable taking values $\pm 1$ with probability $1/2$. Equivalently, under this model, if $E = \sgn(S_0)$, then
\begin{equation} \label{eqn:sign-S-distribution}
E_{ij} = \left\{
\begin{array}{rll}
1, \quad &\mathrm{w.p.} & \rho/2, \\
0, \quad &\mathrm{w.p.} & 1-\rho, \\
-1, \quad &\mathrm{w.p.} & \rho/2.
\end{array}
\right.
\end{equation}
This error model differs from the one assumed in \cite{Candes2009-pp}, in which the error signs come from any fixed (even adversarial) $n \times n$ sign pattern. The stronger assumption that the signs are random is necessary for dense error correction.

Our main result states that under the above assumptions and models, PCP corrects large fractions of errors. In fact, provided the dimension is high enough and the matrix $L_0$ is sufficiently low-rank, $\rho$ can be any constant less than one: 
\begin{thm}[\bf Dense Error Correction via PCP]
Fix any $\rho < 1$. Suppose that $L_0$ is an $n\times n$ matrix of rank $r$ obeying \eqref{eqn:incoherence} with incoherence parameter $\mu$, and the entries of $\mathrm{sign}(S_0)$ are sampled i.i.d. according to \eqref{eqn:sign-S-distribution}. Then as $n$ becomes large\footnote{For $\rho$ closer to one, the dimension $n$ must be larger; formally, $n > n_0(\rho)$. By ``high probability'', we mean with probability at least $1-c n^\beta$ for some fixed $\beta > 0$.}, Principal Component Pursuit \eqref{eqn:pcp} exactly recovers $(L_0,S_0)$ with high probability, provided 
\vspace{-0.08in}
\begin{equation}
\lambda = C_1 \left( 4\sqrt{1-\rho} + \frac{9}{4}  \right)^{-1} \sqrt{\frac{1-\rho}{\rho n}}, \quad
r < \frac{C_2 n}{\mu \log^2 n},
\end{equation}
where $0< C_1 \leq 4/5$ and $C_2 > 0$ are certain constants.
\label{thm:main}
\end{thm}

In other words, provided the rank of a matrix is of the order of $n / \mu \log^2 n$, PCP can recover the matrix exactly even when an arbitrarily large fraction of its entries are corrupted by errors of arbitrary magnitude and the locations of the uncorrupted entries are unknown. 

\noindent{\bf Relations to Existing Results.}
While \cite{Candes2009-pp} has proved that PCP succeeds, with high probability, in recovering $L_0$ and $S_0$ exactly with $\lambda = n^{-1/2}$, the analysis required that the fraction of corrupted entries $\rho$ is small. The new result shows that, with random error signs, PCP succeeds with $\rho$ arbitrarily close to one. This result also suggests using a slightly modified weighting parameter $\lambda$. Although the new $\lambda$ is of the same order as $n^{-1/2}$, we identify a dependence on $\rho$ that is crucial for correctly recovering $L_0$ when $\rho$ is large.

This dense error correction result is not an isolated phenomenon when dealing with high-dimensional highly correlated signals. In a sense, this work is inspired by a conceptually similar result for recovering sparse signal via $\ell_1$ minimization \cite{Wright2008-IT}. To summarize, to recover a sparse signal $x$ from corrupted linear measurements: $y = Ax + e$, one can solve the convex program $\min \|x\|_1 + \|e\|_1, \textup{ s.t. } y = Ax + e$. It has been shown in \cite{Wright2008-IT} that if $A$ is sufficiently coherent and $x$ sufficiently sparse, the convex program can exactly recover $x$ even if the fraction of nonzero entries in $e$ approaches one.

The result is also similar in spirit to results on matrix completion \cite{Candes2008,Candes2009,Gross2009-pp}, which show that under similar incoherence assumptions, low-rank matrices can be recovered from vanishing fractions of their entries.

\section{Main Ideas of the Proof} \label{sec:outline}

The proof of Theorem \ref{thm:main} follows a similar line of arguments presented in \cite{Candes2009-pp}, and is based on the idea of constructing a dual certificate $W$ whose existence certifies the optimality of $(L_0,S_0)$. As in \cite{Candes2009-pp}, the dual certificate is constructed in two parts via a combination of the ``golfing scheme'' of David Gross \cite{Gross2009-pp}, and the method of least squares. However, several details of the construction must be modified to accommodate a large $\rho$. 

Before continuing, we fix some notation. Given the compact SVD of $L_0 = U\Sigma V^*$, we let $T \subset \Re^{n \times n}$ denote the linear subspace $\{UX^* + YV^* \, | \, X,Y\in \R^{n \times r}\}$. By a slight abuse of notation, we also denote by $\Omega$ the linear subspace of matrices whose support is a subset of $\Omega$. We let $\PT$ and $\PO$ denote the projection operators $T$ and $\Omega$, respectively. 

The following lemma introduces a dual vector that in turn, ensures that $(L_0,S_0)$ is the unique optimal solution to \eqref{eqn:pcp}.

\begin{lemma} 
{\bf (Dual Certificate)} Assume $\lambda  < 1-\alpha$ and $\|\P_\Omega \P_T\| \leq 1-\epsilon$ for some $\alpha, \epsilon \in (0,1)$. Then, $(L_0,S_0)$ is the unique solution to \eqref{eqn:pcp} if there is a pair $(W,F)$ obeying 
$$
UV^* + W = \lambda \,(\sgn(S_0) + F + \P_\Omega D)
$$
with $\P_T W = 0$ and $\|W\| \leq \alpha$, $\P_\Omega F = 0$ and $\|F\|_\infty \leq \frac{1}{2}$, and $\|\P_\Omega D\|_F \leq \epsilon^2$.
\label{lem:dual}
\end{lemma}
We prove this lemma in the appendix.  Lemma \ref{lem:dual} generalizes Lemma 2.5 of \cite{Candes2009-pp} as follows:
\begin{enumerate}
\item \cite{Candes2009-pp} assumes that $\|\PO \PT\| \leq 1/2$, whereas we only require that $\|\PO \PT\|$ is bounded away from one. By Lemma \ref{lem:popt}, the former assumption is justified only for small values of $\rho$ (or for small amounts of corruption). 
\item While \cite{Candes2009-pp} requires that $\|W\| \leq 1/2$, we impose a more general bound on $\|W\|$. We find that a value of $\alpha$ closer to 1 gives a better estimate of $\lambda$. 
 \end{enumerate}
For example, by setting $\alpha = 9/10$, to prove that $(L_0,S_0)$ is the unique optimal solution to \eqref{eqn:pcp}, it is sufficient to find a dual vector $W$ satisfying
\begin{equation}
\left \{
\begin{array}{l}
\PT W = 0, \\
\|W\| < \frac{9}{10}, \\
\|\PO (UV^* + W - \lambda \sgn(S_0))\|_F \leq \lambda \epsilon^2, \\
\|\POp (UV^* + W)\|_\infty < \frac{\lambda}{2},
\end{array}
\right .
\label{eqn:prop}
\end{equation}
assuming that $\|\PO \PT\| \leq 1 - \epsilon$ and $\lambda < 1/10$.

We construct a dual certificate in two parts, $W = W^L + W^S$ using a variation of the golfing scheme \cite{Gross2009-pp} presented in \cite{Candes2009-pp}.

\begin{enumerate}
\item {\it Construction of $W^L$ using the golfing scheme.} The golfing scheme writes $\Omega^c = \cup_{j = 1}^{j_0} \Omega_j$, where the $\Omega_j \subseteq [n] \times [n]$ are independent $\text{Ber}(q)$, with $q$ chosen so that $(1-q)^{j_0} = \rho$.\footnote{The value of $j_0$ is specified in Lemma \ref{lem:wl}.} The choice of $q$ ensures that indeed $\Omega \sim \text{Ber}(\rho)$, while the independence of the $\Omega_j$'s allows a simple analysis of the following iterative construction: \vspace{.1in}

Starting with $Y_0 = 0$, we iteratively define
\vspace{-0.08in}
$$
Y_j = Y_{j-1} + q^{-1} \POj \PT(UV^* - Y_{j-1}),
\vspace{-0.08in}
$$
and set 
\vspace{-0.08in}
\begin{equation}
W^L = \PTp Y_{j_0}.
\vspace{-0.08in}
\end{equation}

\item {\it Construction of $W^S$ using least squares.} We set 
\begin{align*}
W^S = \arg \min \| Q \|_F \quad\mathrm{s.t.}\quad \PO Q &= \lambda \, \sgn(S_0), \\ \PT Q &= 0. 
\end{align*}
Since $\| \PO \PT \PO \| = \| \PO \PT \|^2 < 1$, it is not difficult to show that the solution is given by the Neumann series
\vspace{-0.08in}
\begin{equation}
W^S = \lambda \PTp \sum_{k\geq 0} (\PO \PT \PO)^k \sgn(S_0).\label{eqn:ws}
\vspace{-0.08in}
\end{equation}
\end{enumerate}


In the remainder of this section, we present three lemmas that establish the desired main result Theorem \ref{thm:main}. The first lemma validates the principal assumption of Lemma \ref{lem:dual} that $\|\PO \PT\|$ is bounded away from one. The other two lemmas collectively prove that the dual certificate $W = W^L + W^S$ generated by the procedure outlined above satisfies \eqref{eqn:prop} with high probability, and thereby, prove Theorem \ref{thm:main} by virtue of Lemma \ref{lem:dual}.

\begin{lemma}
(Corollary 2.7 in \cite{Candes2009-pp}) Suppose that $\Omega \sim \text{Ber}(\rho)$ and $L_0$ obeys the incoherence model \eqref{eqn:incoherence}. Then, with high probability, $\|\PO \PT\|^2 \leq \rho + \delta$, provided that $1-\rho \geq C_0 \delta^{-2} \frac{\mu r \log n}{n}$ for some numerical constant $C_0 > 0$.
\label{lem:popt}
\end{lemma}

This result plays a key role in establishing the following two bounds on $W^L$ and $W^S$, respectively.

\begin{lemma}
Assume that $\Omega \sim \text{Ber}(\rho)$, and $\|\PO \PT\| \leq \sigma \doteq \sqrt{\rho} + \delta < 1$. Set $j_0 = 2\lceil \log n\rceil$. Then, under the assumptions of Theorem \ref{thm:main}, the matrix $W^L$ obeys, with high probability,
\begin{enumerate}[(a)]
\item $\|W^L\| < 1/10$,
\item $\|\PO (UV^*+W^L)\|_F < \lambda (1-\sigma)^2$,
\item $\|\POp (UV^*+W^L)\|_\infty < \frac{\lambda}{4}$.
\end{enumerate}
\label{lem:wl}
\end{lemma}

The proof of this lemma follows that of Lemma 2.8 of \cite{Candes2009-pp} exactly -- the only difference is that here we need to use tighter constants that hold for larger $n$. The main tools needed are bounds on the operator norm of $\P_{\Omega_j} \P_T$ (which follow from Lemma \ref{lem:popt}), as well as bounds on 
\begin{align*}
\| Q - q^{-1} \P_{\Omega_j} \P_T Q \|_\infty / \| Q \|_\infty, 
 \quad \| Q - q^{-1} \P_{\Omega_j} Q \| / \| Q \|_\infty,
\end{align*}
for any fixed nonzero $Q$ (which are given by Lemmas 3.1 and 3.2 of \cite{Candes2009-pp}). These bounds can be invoked thanks to the independence between the $\Omega_j$'s in the golfing scheme. We omit the details here due to limited space and invite the interested reader to consult \cite{Candes2009-pp}.

%
    
\begin{lemma}
Assume that $\Omega \sim \text{Ber}(\rho)$, and that the signs of $S_0$ are i.i.d.~symmetric (and independent of $\Omega$). Then, under the assumptions of Theorem \ref{thm:main}, the matrix $W^S$ obeys, with high probability, 
\begin{enumerate}[(a)]
 \item $\|W^S\| < 8/10$,
  \item $\|\P_{\Omega^\perp} W^S\|_\infty < \frac{\lambda}{4}$.
\end{enumerate}
\label{lem:ws}
\end{lemma}
See the appendix for the proof details. The proof of this lemma makes heavy use of the randomness in $\mathrm{sgn}(S_0)$, and the fact that these signs are independent of $\Omega$. The idea is to first bound the norm of the linear operator $\mathcal{R} = \PTp \sum_{k\ge 1} (\PO \PT \PO)^k$, and then, conditioning on $\Omega$, we use Hoeffding's inequality to obtain a tail bound for $x^* \mathcal{R}(\mathrm{sgn}(S_0)) y$ for any fixed $x,y$. This extends to a bound on $\| W^S \| = \sup_{\| x \| \le 1,\|y\| \le 1} x^* \mathcal{R}(\mathrm{sgn}(S_0)) y$ via a union bound across an appropriately chosen net. We state this argument formally in the appendix. 

Although the line of argument here is similar to the proof of Lemma 2.9 in \cite{Candes2009-pp}, there are some important differences since that work assumed that $\rho$ (and hence, $\|\PO \PT\|$) is small. Our analysis gives a tighter probabilistic bound for $\| \PTp \sum_{k \geq 1}(\PO\PT\PO)^k E\|$, which in turn yields a better estimate of the weighting parameter $\lambda$ as a function of $\rho$.

\section{Simulations} \label{sec:simulation}
\begin{figure*}[!ht]
 \centerline{\subfigure[$r = 1, C_1 = 0.8$]{\includegraphics[scale=0.19]{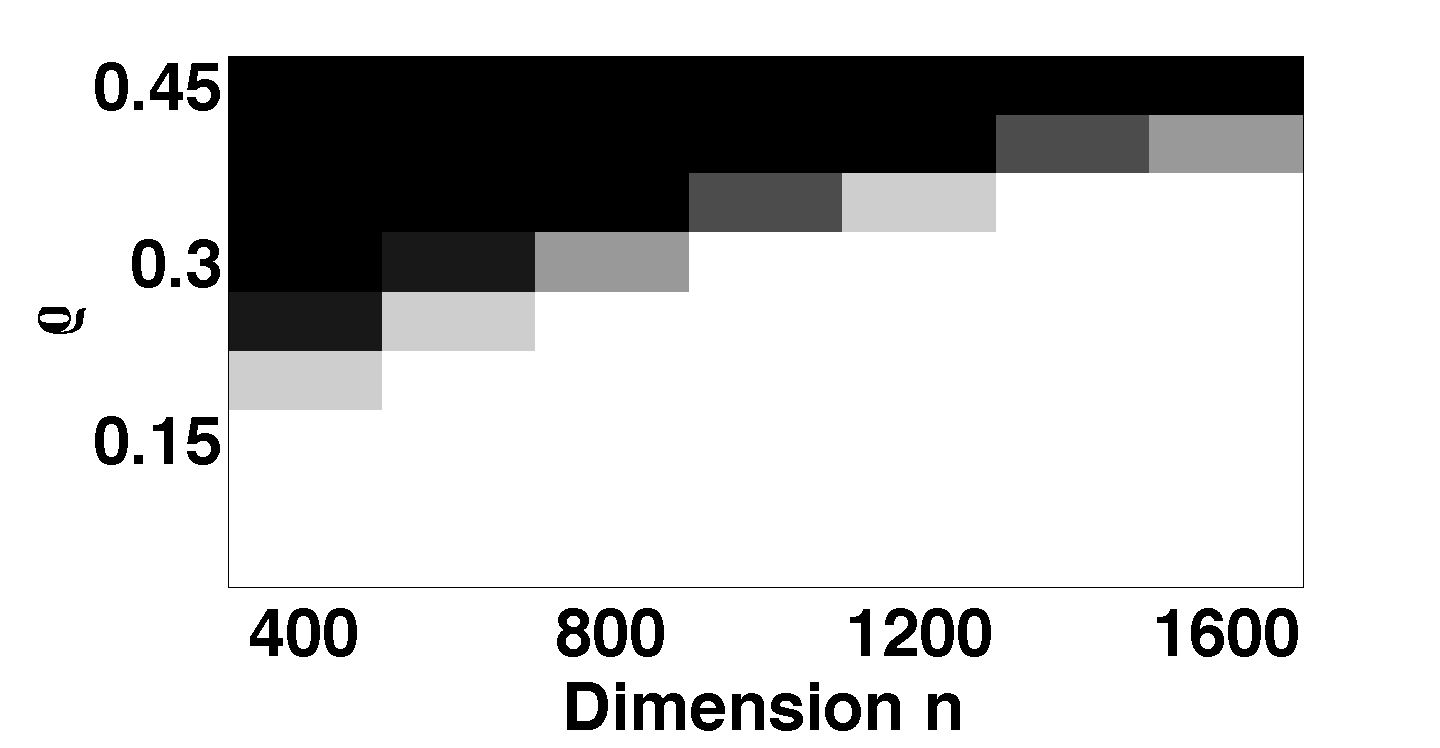} \label{fig:left}} \subfigure[$r=1, C_1 = 4$]{\includegraphics[scale=0.19]{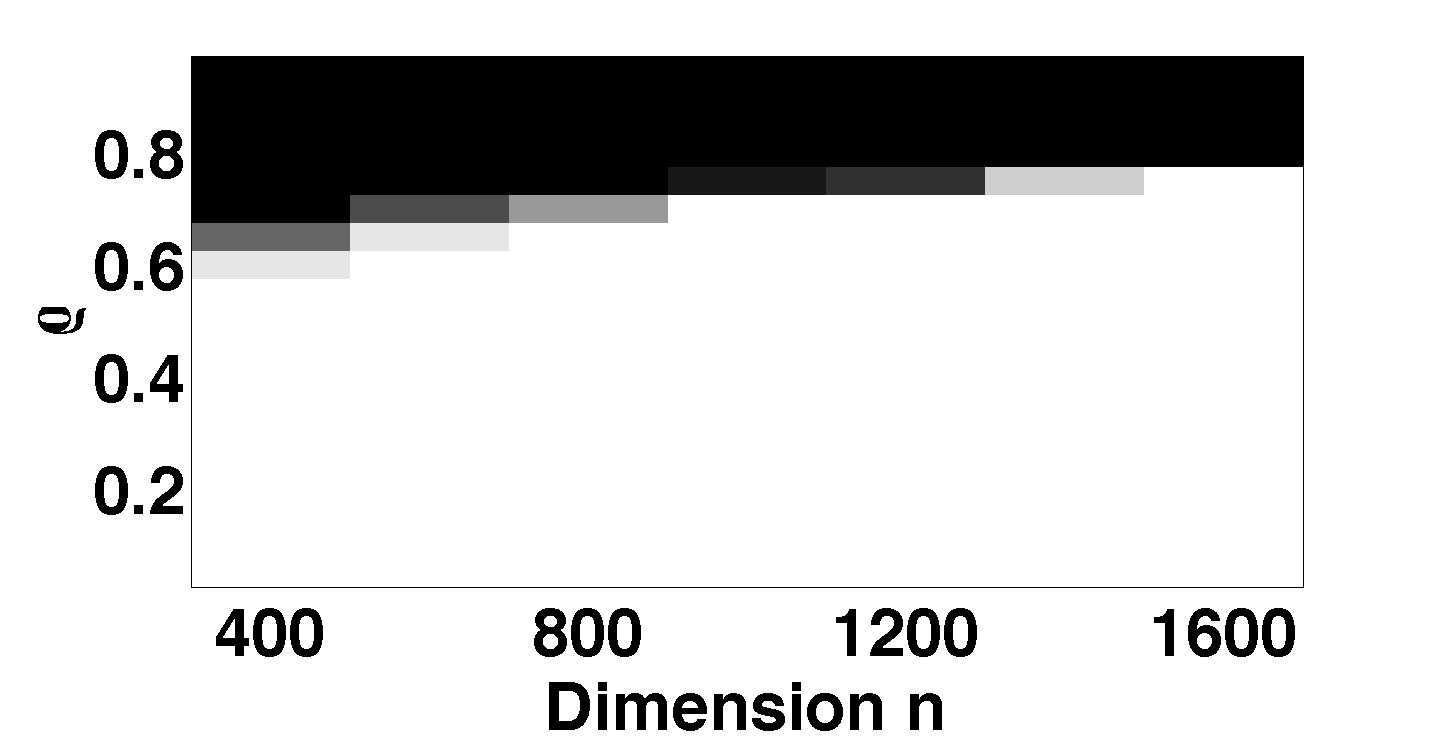} \label{fig:right}} }
  \caption{{\bf Dense error correction for varying dimension.} Given $n$, $r$, and $\rho$, we generate $L_0 = R_1 R_2^*$ as the product of two independent $n \times r$ i.i.d. $\mathcal{N}(0,100/n)$ matrices, and $S_0$ is a sparse matrix with $\rho n^2$ non-zero entries taking values $\pm 1$ with probability $1/2$. For each pair $(n,\rho)$, the plots show the fraction of successful recoveries over a total of 10 independent trials. Here, white denotes reliable recovery in all trials, and black denotes failure in all trials, with a linear scale for intermediate fractions.} \label{fig:sim} 
  \vspace{-0.15in}
  \end{figure*}

In this section, we provide simulation results on randomly generated matrices to support our main result, and suggest potential improvements to the value of $\lambda$ predicted by our analysis in this paper. For a given dimension $n$, rank $r$, and sparsity parameter $\rho$, we generate $L_0$ and $S_0$ as follows:
\begin{enumerate}
\item $L_0 = R_1 R_2^*$, where $R_1,R_2 \in \R^{n \times r}$ are random matrices whose entries are i.i.d.~distributed according to a normal distribution with mean zero and variance $100/n$.
\item $S_0$ is a sparse matrix with exactly $\rho n^2$ non-zero entries, whose support is chosen uniformly at random from all possible supports of size $\rho n^2$.\footnote{As argued in Appendix 7.1 of \cite{Candes2009-pp}, from the perspective of success of the algorithm, this uniform model is essentially equivalent to the Bernoulli model.} The non-zero entries of $S_0$ take value $\pm 1$ with probability $1/2$.
\end{enumerate}

We use the augmented Lagrange multiplier method (ALM) \cite{Lin2009-pp} to solve \eqref{eqn:pcp}. This algorithm exhibits good convergence behavior, and since its iterations each have the same complexity as an SVD, it is scalable to reasonably large matrices. Let $(\hat{L},\hat{S})$ be the optimal solution to \eqref{eqn:pcp}. The recovery is considered successful if $\frac{\|L_0-\hat{L}\|_F}{\|L_0\|_F} < 0.01$, {\it i.e.,} the relative error in the recovered low-rank matrix is less than $1\%$.



For our first experiment, we fix $\mathrm{rank}(L_0) = 1$. This case demonstrates the best possible error correction behavior for any given dimension $n$. We vary $n$ from 400 upto 1600, and for each $n$ consider varying $\rho \in (0,1)$. For each $(n,\rho)$ pair, we choose 
\vspace{-0.1in}
\begin{equation} \label{eqn:lambda}
\lambda = C_1 \cdot \left( 4\sqrt{1-\rho} + \frac{9}{4}  \right)^{-1}\, \sqrt{\frac{1-\rho}{n \rho}}
\vspace{-0.06in}
\end{equation}
 with $C_1 = 0.8$ as suggested by Theorem \ref{thm:main}. Figure \ref{fig:left} plots the fraction of successes across $10$ independent trials. Notice that the amount of corruption that PCP can handle increases monotonically with dimension $n$. 

We have found that the $\lambda$ given by our analysis is actually somewhat pessimistic for moderate $n$ -- better error correction behavior in relatively low dimensions can be observed by choosing $\lambda$ according to \eqref{eqn:lambda}, but with a larger constant $C_1 = 4$. Figure \ref{fig:right} verifies this by repeating the same experiment as in Figure \ref{fig:left}, but with the modified $\lambda$. Indeed, we see larger fractions of error successfully corrected. For instance, we observe that for $n = 1600$, choosing $C_1 = 0.8$ enables reliable recovery when upto $35\%$ of the matrix entries are corrupted, whereas with $C_1 = 4$, PCP can handle upto $75\%$ of corrupted entries. As discussed below, this suggests there is still room for improving our bounds, either by tighter analysis of the current construction or by constructing dual certificates $W^S$ of smaller norm. 

%

 
\section{Discussion} \label{sec:discussion}

This work showed that PCP in fact corrects large fractions of random errors, provided the matrix to be recovered satisfies the incoherence condition and the corruptions are random in both sign and support. The fact that a higher value of the constant $C_1$ offers better error-correction performance in moderate dimensions suggests that the analysis in this work can be further strengthened. In our analysis, the value of $\lambda$ is essentially determined by the spectral norm of $W^S$; it is reasonable to believe that dual certificates of smaller spectral norm can be constructed by methods other than least squares. Finally, while we have stated our results for the case of square matrices, similar results can be obtained for non-square matrices with minimal modification to the proof. 

\vspace{-0.04in}
\section*{Appendix: Proof of Lemma \ref{lem:dual} and Lemma \ref{lem:ws}}

{\bf \noindent Proof of Lemma \ref{lem:dual}.}

\begin{proof}
Let $UV^* + W_0$ be a subgradient of the nuclear norm at $L_0$, and $\sgn(S_0)+F_0$ be a subgradient of the $\ell_1$-norm at $S_0$. For any feasible solution $(L_0+H,S_0-H)$ to \eqref{eqn:pcp}, 
\begin{eqnarray*}
\lefteqn{\|L_0+H\|_* + \lambda\,\|S_0-H\|_1 \; \ge} \\
 &&\|L_0\|_* + \lambda \, \|S_0\|_1 + \langle UV^* + W_0, H\rangle - \lambda \langle \sgn(S_0)+F_0,H\rangle
\end{eqnarray*}
 Choosing $W_0$ such that $\langle W_0, H\rangle = \|\P_{T^\perp}H\|_*$ and $F_0$ such that $\langle F_0, H\rangle = -\|\P_{\Omega^\perp}H\|_1$\footnote{For instance, $F_0 = -\sgn(\POp H)$ and $W_0 = \PTp W$, where $\|W\|=1$ and $\langle W, \PTp H\rangle = \|\PTp H\|_*$. Such a $W$ exists due to the duality between $\|\cdot\|$ and $\|\cdot\|_*$.} gives
 $$
 \begin{array}{l}
\|L_0 + H\|_* + \lambda \|S_0-H\|_1\\
 \geq \|L_0\|_* + \lambda \|S_0\|_1 + \|\P_{T^\perp}H\|_* + \lambda \|\P_{\Omega^\perp}H\|_1 \\ \quad + \langle UV^* - \lambda \sgn(S_0), H\rangle.
 \end{array}
 $$
 By assumption, $UV^*-\lambda \sgn(S_0) = \lambda F - W + \lambda \P_\Omega D$. Since $\|W\| \leq \alpha$, and $\|F\|_\infty \leq \frac{1}{2}$, we have
 $$
 \begin{array}{l}
|\langle UV^*-\lambda \sgn(S_0),H \rangle| \\ 
\leq \alpha \|\P_{T^\perp}H\|_* + \frac{\lambda}{2} \|\P_{\Omega^\perp} H\|_1 + \lambda |\langle \P_\Omega D, H\rangle|.
\end{array}
 $$
 Substituting the above relation, we get
 $$
 \begin{array}{l}
 \|L_0+H\|_* + \lambda\,\|S_0-H\|_1 \\
 \geq \|L_0\|_* + \lambda \|S_0\|_1 + (1-\alpha)\|\P_{T^\perp}H\|_* + \frac{\lambda}{2} \|\P_{\Omega^\perp}H\|_1 \\ \quad - \lambda |\langle \P_\Omega D, H\rangle| \\
 \geq \|L_0\|_* + \lambda \|S_0\|_1 + (1-\alpha)\|\P_{T^\perp}H\|_* + \frac{\lambda}{2} \|\P_{\Omega^\perp}H\|_1\\ \quad - \lambda \epsilon^2 \|P_\Omega H\|_F
 \end{array}
 $$
 We note that
 $$
 \begin{array}{ll}
 \|\P_\Omega H\|_F & \leq \|\P_\Omega \P_T H\|_F + \|\P_\Omega \P_{T^\perp} H\|_F \\
 & \leq (1-\epsilon) \|H\|_F +  \|\P_{T^\perp} H\|_F \\
 & \leq (1-\epsilon) \left (\|\P_\Omega H\|_F +  \|\P_{\Omega^\perp}H\|_F \right ) +  \|\P_{T^\perp} H\|_F
 \end{array}
 $$
 and, therefore,
 $$
 \begin{array}{ll}
  \|\P_\Omega H\|_F & \leq \frac{1-\epsilon}{\epsilon}  \|\P_{\Omega^\perp} H\|_F + \frac{1}{\epsilon}  \|\P_{T^\perp} H\|_F \\
  & \leq \frac{1-\epsilon}{\epsilon}  \|\P_{\Omega^\perp} H\|_1 + \frac{1}{\epsilon}  \|\P_{T^\perp} H\|_*.
  \end{array}
 $$
 In conclusion, we have
 $$
 \begin{array}{l}
  \|L_0+H\|_* + \lambda\,\|S_0-H\|_1 \\
  \geq \|L_0\|_* + \lambda \|S_0\|_1 + \left( (1-\alpha) - \lambda \epsilon \right) \|\P_{T^\perp}H\|_* \\ \quad + \lambda \left( \frac{1}{2} - (1-\epsilon)\epsilon\right) \|\P_{\Omega^\perp}H\|_1.
 \end{array}
 $$
 Because $\| \P_\Omega \P_T \| < 1$, the intersection of $\Omega \cap T = \{ 0 \}$, and hence, for any nonzero $H$, at least one of the above terms involving $H$ is strictly positive.  
 \end{proof}
 
 {\bf \noindent Proof of Lemma \ref{lem:ws}.}
 \begin{proof} 
 
 {\bf Proof of (a).} Let $E = \sgn(S_0)$. By assumption, the distribution of each entry of $E$ is given by \eqref{eqn:sign-S-distribution}. Using \eqref{eqn:ws} we can express $W^S$ as:
 \vspace{-0.08in}
\begin{align*}
 W^S & = \lambda \PTp E + \lambda \PTp \sum_{k \geq 1} (\PO \PT \PO)^k E \\
 & : = \PTp W_0^S + \PTp W_1^S.
\end{align*} 
 For the first term, we have $ \|\PTp W_0^S\| \leq \lambda \|E\|$. Using standard arguments on the norm of a matrix with i.i.d.~entries, we have $\|E\| \leq 4 \sqrt{n \rho}$ with overwhelming probability \cite{Vershynin2007-lec}. 
   
For the second term, we set $\mathcal{R} = \PTp\sum_{k \geq 1}(\PO\PT\PO)^k$, so $W_1^S = \lambda \mathcal{R}(E)$. Notice that whenever $\| \PO \PT \| < 1$, 
\begin{align}
\|\mathcal{R}\| &= \|\PTp\sum_{k\geq 1} (\PO \PT\PO)^k\| \nonumber \\
&\leq \|\PTp \PO \PT \PO\| \cdot \|\sum_{k \geq 0} (\PO \PT \PO)^k \| \nonumber \\
&\leq \|\PTp \PO \PT\| \cdot \|\PT \PO\| \cdot \sum_{k \geq 0} \|\PO \PT \PO\|^k \nonumber \\
&=  \|\PTp \POp \PT\| \cdot \|\PT \PO\| \cdot \sum_{k \geq 0} \|\PO \PT \|^{2k} \nonumber \\
&\leq \frac{\|\POp \PT\| \cdot \|\PO \PT\|}{1 - \|\PT \PO \|^2}. \label{eqn:series}
\end{align}
Consider the two events: 
\vspace{-0.05in}
\begin{align*}
\event_1 &:= \{ \| \P_\Omega \P_T \| \le \sqrt{\rho} + \delta \}, \\
\event_2 &:= \{ \| \P_{\Omega^\perp} \P_T \| \le \sqrt{1-\rho} + \delta \}.
\end{align*}
For any fixed $\eta > 0$, we can choose $\delta(\eta,\rho) > 0$, such that on $\event_1 \cap \event_2$, 
\vspace{-0.06in}
\begin{equation}
\| \mathcal{R} \| \le (1+\eta)\sqrt{\frac{\rho}{1-\rho}}.
\vspace{-.05in}
\end{equation}
Since $\Omega \sim \text{Ber}(\rho)$ and $\Omega^c \sim \text{Ber}(1-\rho)$, by Lemma \ref{lem:popt}, $\event_1 \cap \event_2$ occurs with high probability provided 
\begin{equation}
r \le \delta(\eta,\rho)^2 \min(\rho,1-\rho) n / \mu \log n. \label{eqn:r-bound-4a}
\end{equation}
Since by assumption $r \le C n / \mu \log^2 n$, \eqref{eqn:r-bound-4a} holds for $n$ sufficiently large.

For any $\tau \in (0,1)$, let $N_\tau$ denote an $\tau$-net for $\S^{n-1}$ of size at most $\left(3/\tau\right)^n$ (see \cite{Ledoux} Lemma 3.18). Then, it can be shown that
 $$
 \|\mathcal{R}(E)\| = \sup_{x,y \in \S^{n-1}} \, \langle y, \mathcal{R}(E) x\rangle \leq (1-\tau)^{-2}  \sup_{x,y \in N_\tau} \, \langle y, \mathcal{R}(E) x\rangle
 $$
 For a fixed pair $(x,y) \in N_\tau \times N_\tau$, we define $X(x,y) \doteq  \langle y, \mathcal{R}(E) x\rangle =  \langle \mathcal{R}(yx^*), E\rangle$. Conditional on $\Omega = \supp(E)$, the signs of $E$ are i.i.d. symmetric and by Hoeffding's inequality, we have
 \vspace{-0.1in}
 $$
 \Pr(|X(x,y)| > t | \Omega) \leq 2 \exp \left( -\frac{2t^2}{\|\mathcal{R}(xy^*)\|_F^2}\right).
 \vspace{-0.06in}
 $$
 Since $\|xy^*\|_F = 1$, we have $\|\mathcal{R}(xy^*)\|_F \leq \|\mathcal{R}\|$, so
 $$
 \Pr\left( \sup_{x,y \in N_\tau} |X(x,y)| > t | \Omega \right) \leq 2 |N_\tau|^2 \exp \left( -\frac{2t^2}{\|\mathcal{R}\|^2}\right),
 $$
 and for any fixed $\Omega \in \event_1 \cap \event_2$
 $$
 \Pr\left(\|\mathcal{R}(E)\| > t | \Omega \right) \leq 2 \left(\frac{3}{\tau}\right)^{2n} \exp \left( -\frac{2(1-\tau)^4 (1-\rho) t^2}{(1+\eta)^2 \rho}\right).
 $$
 In particular, for any $C > (1+\eta)\,(1-\tau)^{-2} \sqrt{\log \left( \frac{3}{\tau}\right)}$, $\Omega \in \event_1 \cap \event_2$, $$\Pr\left(\| \mathcal{R}(E)\| > C \sqrt{\tfrac{\rho n}{1-\rho}} \mid \Omega \right) < \exp(-C' n),$$
 where $C'(C) > 0$. Since $\inf_{0 < \tau < 1} \, (1-\tau)^{-2} \sqrt{\log \left( 3/\tau \right)} < 9/4$, by an appropriate choice of $\tau$ and $\eta > 0$, we have
 \begin{equation*}
 \Pr\left( \| \mathcal{R}(E)\| > \frac{9}{4} \sqrt{\frac{\rho n}{1-\rho}} \right) < \exp(-C' n) + \Pr( (\event_1 \cap \event_2)^c ). 
 \end{equation*}
Thus, 
\vspace{-0.15in}
$$
\|W^S\| < \lambda \left( 4\sqrt{\rho} + \frac{9}{4}\,\sqrt{\frac{\rho}{1-\rho}} \right) \sqrt{n} \le 8/10
\vspace{-0.04in}
$$
with high probability, provided $n$ is sufficiently large. 

{\bf Proof of (b)} follows the proof of Lemma 2.9 (b) of \cite{Candes2009-pp}.
\end{proof}

{\small
\bibliographystyle{ieeetran}
\bibliography{dense_error}
}

\end{document}